\documentclass[preprint]{aastex}
\usepackage{portland,lscape}

\def\h2{${}^2$H}

\def\3a{3$\alpha$}
\def\msun{$~M_{\odot}$}
\def\Msun{~M_{\odot}}

\def\lsun{$~L_{\odot}$}
\def\Lsun{~L_{\odot}}

\def\lcno{L$_{\mathrm{CNO}}$}

\def\lH{L$_{\mathrm H}$}

\def\LHe{\mathrm L_{\mathrm{He}}}

\def\gcm{{\mathrm{g\,cm}^{-3}}}
\def\chem#1{${}^{#1}$}

\begin{document}

\title{Structure, Evolution and Nucleosynthesis of Primordial Stars}
 
\author{Lionel Siess\altaffilmark{1,2,3,4}, Mario Livio\altaffilmark{2} and
John Lattanzio\altaffilmark{3}}
\altaffiltext{1}{Institut d'Astronomie et d'Astrophysique CCP 226,
Universit\'e Libre de Bruxelles, B-1050 Bruxelles, Belgium}
\altaffiltext{2}{Space Telescope Science Institute, Baltimore, MD 21218}
\altaffiltext{3}{Department of Mathematics, Monash University, Clayton,
Victoria 3168, Australia}
\altaffiltext{4}{GRAAL, Universit\'e Montpellier II, France} 
 
\authoremail{siess@astro.ulb.ac.be}

\begin{abstract}
 
  The evolution of population III stars ($Z=0$) is followed from the
  pre-main sequence phase up to the AGB phase for intermediate-mass stars
  and up to C ignition in more massive stars. Our grid includes 11 stars,
  covering the mass range 0.8 to 20\msun. During the H and He core
  burning phases an overshooting characterized by $d = 0.20 H_P$ was
  applied and during the AGB phase a small extension of the convective
  envelope was also allowed, characterized by $d = 0.05 H_P$. We find that
  at the beginning of the AGB phase, following the development of a
  convective instability in the He burning shell, a secondary convective
  zone forms at the He-H discontinuity. This unusual convective zone then
  expands and overlaps with the region previously occupied by the receding
  He-driven instability. Carbon in engulfed and a H flash takes place due
  to the activation of the CNO cycle.  Following these successive (H+He)
  flashes, the convective envelope penetrates deeper into the star, reaches
  the secondary H convective shell and allows CNO catalysts to be dredged
  up to the surface. These mixing episodes, which have been found to occur
  in our 1, 1.5, 2, 3, 4 and 5\msun\ models, increase the carbon abundance
  in the envelope and allows low- and intermediate-mass stars to achieve a
  ``standard'' thermally pulsing AGB phase, confirming the recent results
  by Chieffi et al. (2001).  We also find that at the beginning of the
  double shell burning evolution, our 4 and 5\msun\ models experience so
  called ``degenerate'' thermal pulses, which are very similar to those
  found by Frost et al. (1998), but absent from Chieffi et al. (2001)
  simulations. Finally, in the 7\msun\ model, the CNO envelope abundance
  following the second dredge-up is so large that the star does not
  experience the carbon injection episode and follows a standard thermally
  pulsing AGB evolution.  Our computations also indicate that, thanks to
  the small overshooting at the base of the convective envelope, the third
  dredge-up is already operating in stars with $M \ge 1.5$\msun\ after a
  few pulses, and that by the end of our modeling, hot bottom burning is
  activated in stars more massive than $\sim 2$\msun. This evolutionary
  behavior suggests that primordial low- and intermediate stars could have
  been significant contributors to the production of primary \chem{12}C,
  \chem{14}N, and may have contributed to some extent to the production of
  Mg and Al and possibly s-elements (despite the lack of iron seeds) in the
  early universe.
\end{abstract}
 
 
\section{Introduction}

The first stars that formed in our universe, commonly referred to as
Population III stars, have a composition given by the products of the big
bang nucleosynthesis and are essentially made of hydrogen and helium with
some traces of lithium. A major issue concerning primordial stars is the
determination of the initial mass function (IMF). In the absence of heavy
elements and dust grains, the cooling mechanisms may indeed be less
efficient and favor the formation of massive or very massive stars
(e.g. Silk 1977, Bromm et al. 1999). However, it has also been shown
(e.g. Carlberg 1981, Palla et al. 1983) that even a small fraction of
molecular hydrogen can provide a significant contribution to the cooling
due to rotational and vibrational transitions. The resulting Jeans mass of
a pure H and He cloud can then be relatively small and may even fall below
0.1\msun. Yoshii and Saio (1986), using the opacity-limited fragmentation
theory of Silk (1977), estimate the peak mass of the IMF to be around
4-10\msun. Based on one-dimensional hydrodynamical simulations of a
collapsing cloud, Nakamura and Umemura (1999) find that the typical mass of
the first stars is about 3\msun\ which may grow to 16\msun\ by accretion.
Extending their previous work by performing 2D simulations, Nakamura and
Umemura (2001) re-evaluated the IMF and came up with the result that
it is likely to be 
bimodal, with peaks around 1 and 100\msun. Adopting a different approach
based on the effects of a primordial generation of stars on the pollution of
the intergalactic medium, Abia et al. (2001) showed that an IMF peaked
around 4-8\msun\ is needed to account for the large [C,N/Fe]
ratio observed in extremely metal-poor stars. Therefore, the mass scale of
Pop III stars still remains relatively uncertain and it is possible that
even very low-mass stars ($M \la 0.8$\msun) might have survived up to
present day as nuclear burning stars.

In the early universe, low- and intermediate-mass stars actively
contributed to the enrichment of the interstellar medium and an
understanding of their evolution and nucleosynthesis is absolutely
essential for a comprehension of the chemical evolution of the universe. On
the theoretical side, several works have been devoted to the study of the
structure and evolution of primordial stars and we refer the reader to the
recent reviews by Castellani (2000) and Chiosi (2000) for a description of
these previous investigations.

In this paper we present and analyze computations of zero-metallicity
stars, using a state-of-the-art code, thus continuing and completing the
previous studies.  In $\S 2$, we describe briefly the physics of the
stellar evolution code, and in $\S$\ref{evo} we analyze the structure and
evolution of the primordial stars. We also make a careful comparison of our
results with those from previous computations. In $\S$\ref{nucevol}, we
describe the evolution of the surface chemical composition in relation to
the chemical history of the universe and then conclude.

\section{Numerical computations}
\label{num}

The Grenoble stellar evolution code has been used for these
computations. For a detailed discussion of its input physics, we refer the
reader to several papers published by the group (Forestini \& Charbonnel
1997, Siess et al. 2000). Briefly, in the domain of high temperatures ($ >
8000$K) we use the OPAL opacity tables (Iglesias \& Rogers 1996)
which are adequate for enhanced values of C and O. Thus, when
the C and O abundances are modified, i.e. during the third dredge-up
events, their effects on the opacity are taken into account.

The conductive opacities are computed from a modified version of the Iben
(1975) fits to the Hubbard and Lampe (1969) tables for non-relativistic
electrons, from Itoh et al. (1983) and Mitake et al. (1984) for
relativistic electrons and from formulae of Itoh et al.  (1984) as well as
Raikh and Yakovlev (1982) for solid plasmas. Below 8000K, we use the atomic
and molecular opacities provided by Alexander and Fergusson (1994).

The equation of state (EOS) follows the formalism developed by Pols et
al. (1995) and provides an accurate treatment of the non-ideal effects,
especially the treatment of pressure ionization and degeneracy in partially
ionized plasma (see e.g. Siess et al. 2000 for a detailed discussion and
for the tests of this EOS).

The nuclear network includes 180 nuclear (neutron, proton and \chem{4}He
captures) and decay reactions and we follow the abundance evolution of 52
nuclides. Most of the reaction rates which originally came from Caughlan
and Fowler (1988) have recently been updated using the NACRE compilation
(Angulo et al.  1999). Finally, the nuclear screening factors are
parameterized using the Graboske et al. (1973) formalism, including weak,
intermediate and strong screening cases and the computations are
stopped in the massive stars ($M \ga 10\Msun$) at C-ignition when the
temperature reaches $8\times 10^8$K.

The surface boundary conditions are treated in the grey atmosphere
approximation and we use the de Jager mass loss rate prescription (de Jager
et al. 1988).  However, because of the lack of metals, the mass loss rate
only activates during the late stages of the AGB phase.

We use the standard mixing length theory (Cox \& Guili 1968) with
$\alpha_{\rm MLT} = 1.5$ and we follow the Schaller et
al. (1992) suggestion and account for a moderate overshooting
characterized by $d = 0.20 H_P$ during the central H and He burning
phases. During the AGB phase, the extent of the convective envelope has
also been increased by assuming a moderate overshooting
characterized by $d = 0.05 H_P$.  This choice was made as a way of
obtaining a stable Schwarzschild boundary for the envelope in a simple and
consistent way, and is similar to the work of  Herwig et al. (1999).

The chemical composition of our initial models is characterized by $X =
0.765$, $Y = 0.235$ and by a \chem{7}Li mass fraction $X_{\mathrm{Li}} =
9.26\times 10^{-10}$. These values (Burles \& Tytler 1998, Bonifacio \&
Molaro 1997) are taken from Big Bang nucleosynthesis calculations. 
Finally, inside convective zones, the chemical species are homogeneously
mixed.

For numerical details, the code uses an adaptive mesh point. Different
criteria regulate the zoning and in particular, shells are added in regions
where the gradients of energy production and chemical composition are
steep. Typically, the region of mixing episodes is described by $\approx
500$ shells and depending on the evolutionary status, the structure is
discretized into 1000-3500 shells. The time step is constrained so that the
relative variations of the independent variables, between one model to the
next, do not exceed 10\% and the luminosity does not differ by more that
5\%. During the convergence process, the equations of stellar structure
must be satisfied at least to a $10^{-3}$ level of precision.

\section{Structural evolution of low- and intermediate-mass stars}
\label{evo}

In this section, we describe the structure and evolution of low- and
intermediate-mass stars through the computations of 1 and 5\msun\ 
models. These models were chosen so as to enable comparison with other
works.  Firstly, we analyze the evolution during the core H and He burning,
then we describe the AGB phase and finally we summarize the nucleosynthesis
occurring in these stars.

\subsection{Evolution of low-mass stars : 1\msun\ model}
\label{1msun}

During the pre-main sequence (PMS) contraction phase, the small amount of
primordial \chem{7}Li is burnt by proton-captures to helium. After
12.5\,Myr\footnote{The age ``zero'' corresponds to the time we started the
computations, from a polytropic model located on the Hayashi line}, the
central temperature reaches $\simeq 10^7$K and a convective core of $\simeq
0.20$\msun\ forms.  There H is burnt mainly via by the PPI (97\%) chain
which is later supplemented by the PPIII chain. After $3.72\times 10^9$~yr
the convective core disappears, and due to the increased temperature,
carbon production by the \3a reaction starts.  When the central carbon mass
fraction reaches $1.8\times 10^{-12}$ ($T_c = 6.37\times 10^7$K), the CN
cycle ultimately ignites. 

The shifting of the nuclear energy production from the p-p chain to
the CNO cycle takes place roughly at the same carbon abundance threshold.
This is illustrated in Fig. \ref{XC} which
depicts the evolution of the central carbon abundance as a function of
central H abundance. The curves show a bump around a similar
value of $\log X({}^{12}\mathrm{C}) \simeq -11.5$. This indicates an
increase in the H mass fraction, which results from the development of the
convective core, corresponding to the ignition of the CNO reactions. We can
see that stars with $M \ge 3$\msun\ ignite the CNO cycle relatively early
in the evolution, when the central H abundance $X \ge 0.1$. On the other
hand, stars less massive that $\sim 1$\msun\ never ignite CNO reactions
during their main sequence.

The steep temperature dependence of these reactions leads to the
development of a new convective core which lasts for $\approx 16$\,Myr in
our 1\msun\ model. During this ``CNO-flash'', the star exhibits a blue
loop in the HR diagram (HRD) and according to  Fujimoto et al. (1990)
it can be explained as follows : At the ignition of the CN cycle, the
core expands and both 
density and temperature decrease in the central regions.  The surface
layers thus contract and the star moves to the blue side of the HRD. As the
flash decays, core contraction resumes, gravitational energy is released
and a new expansion takes place ; the star moves back to the red side of
the HRD, producing the loop.  When the red giant branch is reached, the
convective envelope deepens and the first dredge up occurs. Note that
only stars with $M \la 1.0\Msun$ experience the first dredge-up. Its effect
on the surface composition is marginal and amounts to a slight enrichment
in \chem{3}He and \chem{4}He and to the destruction of the remaining
\chem{7}Li.

The evolutionary tracks of $Z=0$ stars in the HR diagram is presented in
Fig. \ref{hrd}. The appearance of blue loops in the tracks is present at
the end of the main sequence of our 1\msun\ model and also appears in the
3, 4 and marginally in the 5\msun\ models at the ignition of the CNO cycle
in the HBS at the beginning of the AGB phase.
Note also that due to the lower opacity of the envelope, zero metallicity
stars have a substantially higher effective temperature than their
counterparts of higher metallicity. Their UV flux is substantially larger
and massive primordial stars are now considered to be the leading
candidates for the sources of reionization of the intergalactic medium
(e.g. Gnedin 2000).

The evolution of a $Z=0$ solar mass star has recently been modeled by Weiss
et al. (2000; WCSS hereafter) who compared their work with the earlier
simulation of Fujimoto et al.  (1990; FIH hereafter). We will not reproduce
that comparison here but rather extend it, by comparing our model with the
most recent simulation of WCSS. During the central H burning phase,
the main difference is the development of a convective core in our 1\msun\
model, a feature absent from both the WCSS and the FIH simulations.  The presence
of this convective region may simply arise from the adoption of different
input physics and leads to slightly lower central temperatures and
densities. It also affects the age at the turn off which is slightly larger in
our models (6.56\,Gyr in comparison with 6.31\,Gyr found by WCSS). At the
ignition of the CNO cycle, we find a similar hydrogen concentration at the
center ($X_c = 5.8\times 10^{-4}$), and the maximum size of the CNO-driven
convective core is also very comparable ($\sim 0.11$\msun\ in WCSS, in
comparison with $\sim 0.095$\msun\ in this study). The smaller core
size characterizing our models is probably a consequence of our lower
central temperatures.  For comparison, at the maximum nuclear energy
production during the flash, the central abundances of
\chem{12}C:\chem{14}N:\chem{16}O are $8.30 \times 10^{-12}:2.48 \times
10^{-10}:2.78 \times 10^{-12}$ in our models and $6.50 \times 10^{-12}:2.19
\times 10^{-10}:2.77 \times 10^{-12}$ in WCSS.  Finally, during the
``CNO-flash'' the luminosity generated by the CN cycle (\lcno) remains
always smaller than the one generated by the p-p reactions. At its peak,
\lcno\ represents about 16\% of the total nuclear energy production. The
contribution due to He burning is negligible ($\LHe \simeq 10^{-7}\Lsun$)
and is slightly less than in WCSS ($\LHe = 2.6 \times 10^{-7}\Lsun$) but
still 2 orders of magnitude smaller than in Fujimoto et al.  (1990).

As far as the red giant branch (RGB) evolution is concerned, we did not
encounter the thermal instabilities in the H burning shell (HBS) found by
Fujimoto et al. (1990). Even forcing the timestep not to exceed
$10^{10}$s, we did not find, similarly to WCSS, the instabilities found by
FIH.  Finally, at the time of the He flash, the general properties of our
model are very similar to those of WCSS, the luminosity and
hydrogen-exhausted core mass being 227.75 compared to 229.4\lsun\ and 0.497
compared to 0.492\msun, respectively.


The ignition of core He burning starts off-center at a mass coordinate
$M_r = 0.31\Msun$ which is slightly larger than the 0.25 and 0.27\msun\
found by D'Antona (1982) and Cassisi and Castellani (1993) respectively, but
significantly smaller than the 0.41\msun\ found by FIH. The location where
the first He-shell flash starts has important consequences for the
subsequent evolution. Indeed, if the He-flash ignites close enough to the
H-discontinuity, as has been found by FIH and later by
Fujimoto et al. (2000) and Schlattl et al. (2001), the He-driven convective
zone might be able to penetrate the overlaying H rich layers. In this
situation, protons are engulfed in the He-driven convective zone, mixed
with the abundant carbon and a H-flash occurs which subsequently leads to
large modifications in the surface chemical composition and in particular
to a large C and N enrichment. The location where the He flash ignites is
thus a crucial issue but it depends on the star's internal structure, which
in turn reflects the imprint of the adopted physics. 
Recently, Schlattl et al. (2001) have analyzed in detail the conditions for
proton ingestion during the He-core flash. Their analysis reveals that the
occurrence of this phenomenon depends quite significantly on the initial
conditions. More specifically, they showed that the use of a different EOS
(which determines the electron degeneracy), of different radiative and/or
conductive opacity tables, or the treatment or not of diffusion, can
trigger or inhibit the penetration of the He-driven flash into the H rich
layers. They also showed that the treatment of convection has very little
effect on the outcome. As emphasized in FIH, the
use of different treatments of plasma neutrinos turn out to have
significant effects on the results. Finally, we recall that  the location
of the He flash also depends on the rate at which the H burning shell
advances in mass. Since there are several differences in physical input
between FIH and the present work (which include core overshooting),  it is
not easy to give a unique explanation for 
this different location, but we suspect that the opacities may have played
a dominant role. To conclude, we didn't find proton mixing because our He
ignites too far from the H/He discontinuity and to illustrate the entangled
dependence of this phenomenon on the initial conditions, let us note that
Marigo et al. (2001) did not mention any evidence of extra protons mixing
during the He core flash of their low-mass stars.

The flash is initiated at the point of maximum temperature.
Where the degeneracy is lifted, the temperature increases very rapidly and
the nuclear energy production due to \3a reactions undergoes a runaway. A
convective instability develops and the stellar structure expands.  The
pressure and the temperature in the degenerate core decrease as the weight
of the overlaying layers decreases and when the flash dies away, core
contraction resumes. The temperature peak then moves inward and advances in
mass by $\sim 0.03$\msun\ between each secondary flash, a value very
similar to the one found by Mengel and Sweigart (1981) in their 0.70\msun\
stellar model.

Figure \ref{flash} illustrates the sequence of events that precedes central
He burning. A series of 10 convective instabilities progressively lift the
degeneracy in the core. The first flash is the strongest due to the higher
core degeneracy at that time and has the largest extent. The extension in
mass of the subsequent flashes is smaller due to the lower He-luminosity
and the time interval between the secondary flashes also gradually
increases (see also Despain 1981). Table \ref{tab1} summarizes the
characteristics of these flashes.
\begin{deluxetable}{cccc} 
\tablecolumns{4} 
\tablewidth{0pc} 
\tablecaption{Characteristics of the He flashes occurring before
  central He burning}
\tablehead{pulse \# & $L_{\rm nuc}$ (\lsun) & $M_{\rm ignit}$
\tablenotemark{a} & duration (yr)}
\startdata
1 & $7\times 10^9$   &    0.2991 & 2300 \\
2 & $1.4\times 10^5$ &    0.2655 & 1140 \\
3 & $7.9\times 10^4$ &    0.2487 & 1700 \\
4 & $5.3\times 10^4$ &    0.2289 & 2600 \\
5 & $2.3\times 10^4$ &    0.1938 & 3400 \\
6 & $1.4\times 10^4$ &    0.1740 & 4500 \\
7 & 6000             &    0.1420 & 7900 \\
8 & 2330             &    0.1145 & 16900 \\
9 & 1020             &    0.0886 & 31500 \\
10 & 400             &    0.0672 & 87000 \\
\enddata
\tablenotetext{a}{mass coordinate of the inner border of the convective
  zone at the ignition of the instability (in\,\msun)} 
\label{tab1}
\end{deluxetable}

During the major flash, only a very small amount of protons is ingested
into the convective region and this prevents the H-burning luminosity from
reaching the very high values encountered in the computation of Hollowell
et al. (1990) and Fujimoto et al. (2001). Our computations shows that the H
mass fraction at the outermost point of the convective region is always
smaller than $X< 10^{-8}$ and the hydrogen-generated luminosity never
exceeds $\sim 10^4$\lsun\ at its peak. The maximum extent of the convective
flash reaches to within 0.005\msun\ of the H rich region (where $X > 0.1$)
which is located at $M_r = 0.5075$\msun.

After 10 flashes, He burning starts at the center and the star
rapidly moves to the blue side of the HRD along the horizontal branch. The
envelope recedes and a convective core of $\sim 0.20\Msun$ forms and is
sustained for $\approx 70.6$\,Myr until He is depleted at the center. The
energy production comes largely from the \3a reactions (90\%) taking place
in the core but also in part from gravitational contraction (10\%). The
HBS, while still active, does not contribute significantly to the net
energy budget.  During the central He-burning phase, the convective core is
subject to ``breathing'' pulses and convective tongues develop at the
surface of the core adding additional He into the burning region and
temporarily increasing the He-burning luminosity. 

We present in Tab. \ref{tab3} selected parameters for our computed
models during the central H and He burning phases, namely, the duration of
H (He) core burning phase, the central temperature and density when the
central abundance of H (He) equal 0.5, the abundance of the 3 most abundant
species at the end of each core burning phase, the core mass at the tip of
the RGB and the deepest extent of the convective envelope following the He
core burning phase.  Interestingly, we see that the central density during
the H core burning and the core mass at the tip of the RGB show a minimum
around $M \simeq 1.5$\msun. This value is very similar to the one found by
Cassisi and Castellani (1993) and corresponds to the critical mass for the
ignition of He in strongly degenerate conditions. This smaller value for
the mass limit, compared for example with the 1.8\msun\ found for
$Z=10^{-6}$ (Cassisi \& Castellani 1993), results from the fact that more
metal deficient stars must increase their temperature (and density) to
sustain a larger luminosity. They are consequently hotter and are thus less
affected by electron degeneracy.
\begin{deluxetable}{cccccccc} 
\tabletypesize{\scriptsize}
\tablecolumns{8} 
\tablewidth{0pc} 
\tablecaption{Evolutionary characteristics of $Z=0$ stars}

\tablehead{
\multicolumn{8}{c}{H burning phase} \\
\cline{1-8} \\
\colhead{} & \colhead{} & \multicolumn{2}{c}{$X=0.5$ at the center} &
\multicolumn{3}{c}{$X< 10^{-8}$ at the center} & \colhead{} \\
\cline{3-4} \cline{5-7} \\
Mass  & ${t_{\rm H}}$\tablenotemark{a} & ${T_c}$ ($10^6$K) & $\log
\rho_c$ & \chem{12}C & \chem{14}N & \chem{16}O &
${M_{\rm tip}}$\tablenotemark{b}}

\startdata

0.8   & 1.48e10 & 13.9 & 2.161 & 2.03 (-10) & 3.50 (-10) & 1.68 (-11) & 0.468 \tablenotemark{c} \\
1.0   &   6.85e9  &  16.7   & 2.169 & 7.71 (-10) & 1.71 (-9) & 1.59 (-10) & 0.494 \\
1.5   &   1.89e9  &  23.2   & 2.050 & 8.73 (-10) & 3.07 (-9) & 2.87 (-10) & 0.277  \\
2.0   &   7.42e8  &  27.8   & 2.173 & 8.68 (-10) & 7.18 (-9) & 5.67 (-10) & 0.334 \\
3.0   &   2.12e8  &  38.9   & 2.267 & 7.12 (-9) & 2.27 (-8) & 2.35 (-9) & 0.445 \\
4.0   &   1.04e8  &  49.1   & 2.303 & 9.17 (-8) & 1.81 (-8) & 1.50 (-8) & 0.516 \\
5.0   &   6.14e7  &  58.8   & 2.338 & 1.74 (-6) & 1.64 (-7) & 2.15 (-8) & 0.570 \\
7.0   &   3.16e7  &  75.0   & 2.385 & 8.47 (-6) & 3.05 (-7) & 4.78 (-8) & 0.900 \\
10    &   1.82e7  &  91.7   & 2.398 & 1.16 (-5) & 3.88 (-7) & 5.79 (-8) & 1.491 \\
15    &   1.21e7  &  100.0  & 2.152 & 3.37 (-5) & 4.86 (-7) & 9.39 (-8) & 3.177 \\
20    &   8.77e6  &  103.7  & 2.022 & 1.85 (-4) & 6.27 (-7) & 2.66 (-7) & 5.196 \\

\cutinhead{He burning phase}
\colhead{} & \colhead{}  &  \multicolumn{2}{c}{$Y=0.5$ at the center}
& \multicolumn{3}{c}{$Y < 10^{-8}$ at the center} & \colhead{} \\
\cline{3-4} \cline{5-7} \\
Mass & ${t_{\rm He}}$\tablenotemark{d}& $T_c$ ($10^8$K) & $\log \rho_c$ &
 \chem{12}C & \chem{16}O & \chem{20}Ne & ${M_{\rm env}}$\tablenotemark{f} \\
\cline{1-8} \\
1.0   &  175.6 & 1.305 &  4.113 & 0.294 & 0.698 & 7.94 (-3) \tablenotemark{e} & 0.602 \\
1.5   &  104.8 & 1.291 &  4.146 & 0.194 & 0.801 & 4.45 (-3) \tablenotemark{e} & 0.577\\
2.0   &  68.3  & 1.340 &  4.010 & 0.259 & 0.741 & 1.00 (-5) & 0.690\\
3.0   &  17.7  & 1.430 &  3.766 & 0.378 & 0.622 & 3.69 (-6) & 0.756 \\
4.0   &  12.0  & 1.504 &  3.620 & 0.424 & 0.576 & 3.62 (-6) & 0.842\\
5.0   &  8.8   & 1.539 &  3.506 & 0.372 & 0.627 & 6.45 (-6) & 0.898\\
7.0   &  4.0   & 1.640 &  3.300 & 0.380 & 0.620 & 1.41 (-5) & 1.054\\
10    &  3.0   & 1.699 &  3.132 & 0.355 & 0.645 & 8.30 (-5) & 2.380\\
15    &  0.90  & 1.877 &  2.921 & 0.296 & 0.704 & 4.78 (-4) & no DUP\\
20    &  0.63  & 1.965 &  2.818 & 0.280 & 0.718 & 1.06 (-3) & no DUP\\

\enddata

\tablenotetext{a}{main sequence duration}
\tablenotetext{b}{mass of the H depleted core at the time of He ignition as defined when
$\LHe > 100 \Lsun$ (in\,\msun)}
\tablenotetext{c}{At the end of our computations (15\,Gyr), this star had
not yet ignited its core He supply. The value quoted is for the last
computed model.} 
\tablenotetext{d}{duration of core He burning phase (Myr)}
\tablenotetext{e}{in this model, the third most abundant element is
\chem{22}Ne}
\tablenotetext{f}{mass coordinate of the maximum extent of the convective
envelope following the He core burning phase in \msun}
\label{tab3}
\end{deluxetable}

The comparison of the nuclear lifetime with the recent paper by Marigo et
al. (2001) reveals discrepancies as large as 15\% in the duration of the
main sequence. The largest shift is found for the critical mass
$M=1.5$\msun, and for this reason we suspect that part of these differences
results from the adopted equation of state. Further, these authors use a
different value for the overshooting parameter and this can also explain
the discrepancies. Globally, our lower mass models ($M\le 3$\msun) evolve
more slowly while the more massive ones tend to evolve slightly faster.

At the end of the central He burning phase ($t \sim 7.27\times 10^9$~yr), a
CO core of $\sim 0.202\Msun$ has formed and the star enters the early AGB
phase. As core contraction proceeds, the luminosity of the He burning
shells increases and the H burning shell is switched-off. This
situation allows the deepening of the convective envelope which reaches a
mass coordinate of 0.599\msun. The second dredge-up is a general feature of
stars in the mass range $1.0 \le M < 10$\msun\ and its efficiency in 
polluting the envelope increases with mass. Following this deep mixing, the
surface layers are mainly enriched in \chem{4}He and to a lesser extent
with CNO elements. Table \ref{tab2} summarizes some of the properties of
the envelope composition at the end of this phase.

\subsection{Evolution of intermediate-mass stars : 5\msun\ model}

The lack of \chem{12}C prevents the CN cycle from operating and during the
major phase of central H burning, the dominant source of energy of the star
is provided by the p-p chains.  Because of the relatively weak temperature
sensitivity and low efficiency of these reactions, the active burning
region ($\varepsilon_{\mathrm{nuc}} > 5\,$erg\,g$^{-1}$s$^{-1}$)
encompasses a large fraction of the star, up to $\simeq 80\%$ of its
mass. At the beginning of the main sequence, a convective core develops and
disappears at an age of 41.6\,Myr when the central H abundance has been
reduced to $X_c \simeq 0.34$. This value is slightly smaller than the one
reported by Chieffi and Tornamb\'e (1984) who found $X_c \simeq 0.45$ and
the difference can be attributed to the use of different nuclear reaction
rates in the earlier work. Indeed, the nuclear reaction rates for the p-p
reactions provided by the NACRE compilation (Angulo et al. 1999) are either
identical or smaller than the ones used by Chieffi and Tornamb\'e. As a
consequence of these slower rates, the temperature must be higher in order
to increase the nuclear energy production and balance the stellar energy
budget. This explains why our 5\msun\ model is globally hotter and evolves
more rapidly than in the computations of Chieffi and Tornamb\'e. The
differences in the opacity tables also contribute to increase the
discrepancies as illustrated by the duration of the main sequence, which is
35\% shorter in our simulations (62.6\,Myr in our work compared to
100\,Myr). On the other hand, comparisons with the more recent
computations of Chieffi et al. (2001, hereafter CDLS) show a good agreement
on the main sequence lifetime (62\,Myr in their simulations) and a close
examination of their Fig.~2 reveals that during the main core burning
phases, the evolution of the central density and temperature are very
similar. However, the duration of the first convective core phase and the
central H abundance when it disappears are significantly different (they found
20\,Myr and $X_c = 0.5$). Since the input physics of both stellar evolution
codes are similar (same nuclear reaction rates, opacity tables) and since we
do not expect large differences in the EOS at this evolutionary phase, we
attribute this discrepancy as probably due to the presence of overshooting
in the present
work.  During the main sequence, \chem{12}C slowly builds up in the core
and when its central abundance reaches $\simeq 5\times 10^{-12}$, the CNO
cycle comes into play and rapidly takes over the nuclear energy
production. Due to the reaction's strong temperature dependence, a central
convective zone forms again and extends up to $\sim 0.56$\msun\
(Fig. \ref{kippen5}), a value substantially larger than the one found
by CDLS (0.4\msun). Note that during this ``convective'' period, \3a
reactions keep running. They participate only marginally to the nuclear
energy production but maintain and increase the central \chem{12}C
abundance.  When the central H mass fraction falls below $5\times 10^{-6}$,
the convective core disappears and a H burning shell (HBS)
forms. Contraction resumes, and after 2.4\,Myr a helium convective core
develops while the star is still located in the blue side of the HRD. This
failure to become a red giant just after central hydrogen depletion
prevents the first dredge-up from occurring.  During the He burning phase,
the central temperature increases and when it reaches $T_c \simeq 1.7\times
10^8$K, carbon starts to be burnt into \chem{16}O by $\alpha$ captures. At
the end of the central He burning phase, a degenerate core of $\sim 0.6$\msun\ has
formed, composed of 37\% of \chem{12}C and 63\% of \chem{16}O. Note
that the 10\,Myr duration of the He burning phase is very comparable to the
9.8\,Myr found by CDLS.

When He is depleted at the center, the star undergoes a large expansion and
moves rapidly to the red side of the HRD. The HBS quenches and the
convective envelope deepens. At its largest extent, the envelope reaches a
mass coordinate of $M_r \simeq 0.896\Msun$ and dredges up to the surface
the products of the nucleosynthesis taking place in the H burning. As
a result of the CNO cycle, the surface is mainly enriched in \chem{14}N and
\chem{4}He and because the \3a reactions are also at work in the HBS, the
abundance of \chem{12}C and \chem{16}O are increased while H and \chem{3}He
are depleted. After the reignition of the HBS, the star undergoes a series
of ``weak'' thermal runaways in the HeBS ($\LHe \la 10^5$\lsun) which are
not powerful enough to trigger the formation of a convective region, nor to
extinguish the HBS (Fig. \ref{kippen5}). However, at the 13$^{th}$
instability, a convective region forms in the HeBS, and by the mechanism
described in the next section, a pollution of the envelope takes place,
allowing the star to resume a ``standard'' AGB evolution.  Interestingly,
Chieffi and Tornamb\'e (1984) also found that 4 small instabilities
developed in their 5\msun\ model at the beginning of the AGB
phase. However, these authors did not find subsequent mixing and  a
steady state double shell burning evolution followed, characterized by the
absence of thermal pulses.

After the H reignition, the luminosity is mostly provided by the HBS (75\%)
and to a lesser extent by the HeBS (15\%) and gravitational contraction
(10\%).  When the temperature in the HBS reaches $10^8$K, the \3a reactions
start to operate more efficiently and \chem{12}C production
increases. However, due to the relatively low density in the HBS
($\rho_{\mathrm{HBS}} \la\ 350 \gcm$), the \3a reactions do not contribute
significantly to the nuclear energy generation, which is mainly the result
of efficient CNO burning. The high temperatures found in the HBS also
activate additional CN cycling. In particular, the chain of reactions
\chem{16}O(p,$\gamma$)\chem{17}F($\beta^+$)\chem{17}O(p,$\gamma$)\chem{18}F($\beta^+$)\chem{18}O(p,$\gamma$)\chem{19}F
is responsible for some marginal \chem{19}F production. However, the 
main surface enrichment in this element is  the result of the subsequent
deep third dredge-up (3DUP) episodes.  By the end of our computations,
its surface mass fraction amounts to $X(^{19}\mathrm F) \simeq 6.4\times
10^{-8}$, but as for \chem{7}Li (see Sect. \ref{nucevol}), this value depends
on the 
competitive effects resulting from its  destruction by hot bottom burning and
enrichment by the third dredge-up events. Note also that
at the base of the HBS, the ``hot'' CNO cycle is operating, characterized
by the main branching
\chem{13}N(p,$\gamma$)\chem{14}O($\beta^+$)\chem{14}N.  The NeNa and MgAl
cycles are also activated and contribute locally to the production of
\chem{23}Na and \chem{26}Al, but the abundances of these elements still
remain very small ($< 10^{-15}$ in mass fraction).


\subsection{The Thermally pulsing AGB phase}

\subsubsection{The Carbon Injection}
\label{mixepisode}

Thermal pulses during the AGB evolution are present in all of our
models from 1.0 to 7\msun. The occurrence of
thermal pulses in zero-metal stars was initially investigated theoretically
by Fujimoto et al. (1984). These authors first pointed out the existence of
threshold values for the core-mass and CNO abundances for the occurrence of
thermal instabilities. In particular, they showed that if the core mass is
smaller then a critical value $M^* \approx 0.73$\msun, recurrent He shell
flashes are expected to occur. They also found that for $M_{\rm core} \ga
M^*$, there exists a critical envelope CNO abundance ($\sim 10^{-7}$) above
which He shell flashes will be triggered. The reason for this behavior
was explained in Fujimoto et al. (1984) and Chieffi and Tornamb\'e (1984)
and can be summarized as follows : in massive stars, for which $M_{\rm
core} \ga M^*$, the lack of CNO nuclei imposes that H is burnt at very high
temperatures, allowing for carbon production in the HBS. Thus, the \3a
reactions are working contemporaneously in both shells and they advance in
mass at a similar rate. This prevents the mass growth of the intershell and
thereby the occurrence of the instability. Conversely, in low-mass stars,
the energy requirements are lower and the \3a reactions are not very active
in the HBS. The two shells thus proceed outward at different rates - an
unstable situation leading to the development of thermal pulses.
\begin{deluxetable}{ccccccc} 
\tabletypesize{\scriptsize}
\tablecolumns{7}
\tablewidth{0pc} 
\tablecaption{Selected properties of our thermally pulsing stars}
\tablehead{
\colhead{}    & \colhead{}    &  \multicolumn{5}{c}{Surface mass fraction} \\
\cline{3-7} \\
\colhead{Mass}  & \colhead{$M_{\rm core}$\tablenotemark{a}} & \colhead{H} &
\colhead{\chem{4}He} & \colhead{\chem{12}C} & \colhead{\chem{14}N} &
\colhead{\chem{16}O}}  
\startdata

\cutinhead{Prior to first pulse} \\
1.0 & 0.4176 & 0.7603 & 0.2389 & 2.67  (-32) & 3.46  (-28) & 3.18  (-26) \\
1.5 & 0.4979 & 0.7300 & 0.2697 & 7.57  (-22) & 4.57  (-22) & 1.22  (-21) \\
2.0 & 0.6476 & 0.6924 & 0.3074 & 2.71  (-20) & 8.43  (-21) & 2.84  (-21) \\
3.0 & 0.6890 & 0.6421 & 0.3578 & 9.63  (-17) & 1.09  (-14) & 3.06  (-16) \\
4.0 & 0.7733 & 0.6348 & 0.3652 & 8.92  (-14) & 8.99  (-12) & 2.27  (-13) \\
5.0 & 0.8833 & 0.6205 & 0.3795 & 6.873 (-9)  & 7.72  (-10) & 1.18 (-10) \\
7.0 & 1.0268 & 0.6236 & 0.3764 & 2.409 (-5)  & 1.444 (-9)  & 9.390 (-8) \\\\

\cutinhead{End of carbon injection} \\

1.0 & 0.4968 & 0.7544 & 0.2447 & 9.733 (-6) & 7.596 (-5) & 1.141 (-4) \\
1.5 & 0.5041 & 0.7117 & 0.2879 & 4.188 (-7) & 1.739 (-5) & 2.425 (-7) \\
2.0 & 0.6662 & 0.6175 & 0.3081 & 1.475 (-7) & 5.730 (-6) & 5.760 (-8) \\
3.0\tablenotemark{b} & 0.7381 & 0.6421 & 0.3578 & 2.65 (-11) & 6.91 (-10) & 5.59 (-12) \\
3.0 & 0.7413 & 0.6421 & 0.3578 & 5.69 (-11) & 2.247 (-9)  & 2.29 (-11) \\
3.0 & 0.7464 & 0.6421 & 0.3578 & 4.95 (-10) & 1.814 (-8) & 1.91 (-10) \\
3.0 & 0.7487 & 0.6421 & 0.3578 & 2.193 (-7) & 9.312 (-8) & 1.846 (-6) \\
4.0 & 0.8316 & 0.6342 & 0.3655 & 5.043 (-6) & 1.527 (-6) & 1.419 (-4) \\
5.0 & 0.8971 & 0.6204 & 0.3795 & 1.163 (-5) & 2.071 (-7) & 1.524 (-5) \\

\cutinhead{Last computed model}

1.0 & 0.627 & 0.7526 & 0.2466 & 7.473 (-4) & 7.854 (-5) & 2.231 (-4)  \\
1.5 & 0.615 & 0.6882 & 0.3019 & 6.905 (-3) & 1.922 (-5) & 2.356 (-3)  \\
2.0 & 0.739 & 0.6811 & 0.3153 & 3.376 (-3) & 6.718 (-6) & 8.515 (-5)  \\
3.0 & 0.812 & 0.6312 & 0.3646 & 4.447 (-4) & 3.256 (-3) & 1.121 (-4)  \\
4.0 & 0.849 & 0.6312 & 0.3660 & 4.453 (-6) & 1.321 (-3) & 1.167 (-3)  \\
5.0 & 0.908 & 0.6187 & 0.3809 & 6.138 (-5) & 2.963 (-4) & 2.317 (-5)  \\
7.0 & 1.059 & 0.6226 & 0.3773 & 8.939 (-6) & 7.838 (-5) & 1.750 (-6)  \\

\enddata 

\tablenotetext{a}{mass at the base of the He burning shell (in\,\msun)}
\tablenotetext{b}{this model experiences 4 carbon injections before reaching
the TP-AGB phase}
\label{tab2}
\end{deluxetable}

Our simulations indicate that contrary to the theoretical expectations of
Fujimoto et al. (1984),
stars with $M_{\rm core} > 0.73$\msun\ do  enter the thermally pulsing
AGB (TP-AGB) phase. The reason for this behavior can be found in the fact
that, low- and intermediate-mass stars initially experience a series of
instabilities (carbon injections) that progressively enrich the HBS and the
envelope in heavy 
elements. As a consequence of this chemical ``pollution'', the CNO cycle
operates in the HBS without the requirement of carbon production by \3a
reactions, as in more metal rich stars. The sequence of events leading to
this carbon injection can be illustrated through a careful examination of our
3\msun\ model.

At the reignition of the HBS, the two burning shells are not advancing in
mass at the same rate. The intershell mass increases and the He burning
shell is subject to small-amplitude thermal pulses. When the convective 
instability develops in the HeBS, it rapidly grows in mass and reaches the
tail of the HBS, in a region where the H mass fraction is $X \la 10^{-4} - 10^{-5}$ and
it is stopped by the large entropy barrier caused by the HBS (Iben
1976, Fujimoto 1977). As the He convective shell (HeCS) retreats, a H
convective shell (HCS) forms at the H-He interface (Fig. \ref{m3ingest} and
\ref{m2ingest}). The subsequent expansion of the HCS  in the
underlaying layers previously occupied by the pulse dredges up
some carbon and initiates a H
flash. During this secondary flash, the luminosity generated by the HCS
(\lH) reaches a maximum value of 
$\sim 3\times 10^6$\lsun\ and greatly exceeds the He-generated
luminosity. The nuclear energy production in the HCS increases very rapidly
because the shell is confined to a very narrow mass range ($\Delta
m_{\mathrm{HCS}} \simeq 2\times 10^{-5}$\msun) where the temperature is on
the average high for proton burning ($T_{\mathrm{HCS}} \sim 95 \times
10^6$K). As we mentioned earlier in the case of the 5\msun\ model, the
luminosity of the HBS is mainly provided by the CNO cycle. However, in this
model, the shell has a very low C abundance ($\sim 10^{-8}$) and the sudden
ingestion of a large amount of carbon inevitably boost the nuclear
energy production.

The energy released during the successive (He+H) flashes produces an
expansion of the overlaying layers which is followed by a deeper
penetration of the convective envelope. In its descent, the envelope 
penetrates the region that was previously occupied by the HCS and enriches 
the surface with the elements present in the HCS. During this event,
the envelope abundances of \chem{12}C, \chem{14}N and \chem{16}O increase
significantly (see Tab. \ref{tab2}).

This irreversible process, which we call ``carbon injection'', is 
characteristic of all our intermediate mass
models, even in models with $M_{\rm core} \la M^*$ and is activated only
during the first thermal pulse of our 1, 1.5, 2, 4 and 5\msun\ models and
during the first 4 instabilities in our 3\msun\ model
(Fig. \ref{m3ingest}).  In this latter case, the first carbon injections
are not able to raise the CNO envelope abundance above the threshold value
of $\sim 10^{-7}$ and the process repeats until this value is
exceeded. Table \ref{tab2} summarizes some properties at the end of these
carbon injections. We also note that in our 7\msun\ model, the carbon
enrichment of the envelope following the second dredge-up is large enough
to ensure that the CNO cycle is operating efficiently inside the HBS
without the need of self carbon production by \3a reactions. Therefore,
this model reaches a ``standard'' AGB phase and the C-injection are
absent. CDLS found that their 6\msun\ model also ended the early AGB phase
with a large CNO abundance and thus, during the following evolution, it
behaved normally, like our 7\msun\ model. As far 
as our computations are concerned, carbon does not ignite in this model, a
conclusion also reached by CDLS. This result thus indicates that the
minimum mass for degenerate carbon ignition is $M_{\rm up} \ge 7\Msun$.
We present in Tab. \ref{tabm7} and \ref{tabm3}  some properties for the 7
and 3\msun\ models.  Columns
(1)-(13) list the following quantities (1) the location of the H-burning 
shell at the pulse ignition, (2) the location of the convective envelope
relative to the HBS at the pulse ignition ($M_{\rm env} = M_H+\Delta M_{\rm
env}^{\rm H}$), (3) the pulse duration ($\Delta t_{\rm pul}$), (4) the
interpulse duration ($\Delta t_{\rm inter}$ in $10^3$~yr), the temperature
(5) and density (6) at the base of the pulse at the time of maximum nuclear
energy generation, (7) the peak luminosity of the pulse, (8) the location
of the top of the convective pulse relative to the HBS ($M_{\rm pul}^{\rm
top} = M_H-\Delta M_{\rm pul}^{\rm H}$), (9) the pulse mass ($M_{\rm pul} =
M_{\rm pul}^{\rm top}-M_{\rm pul}^{\rm bot}$), (10) the location of the
deepest penetration of the convective envelope relative to HBS after the
pulse ($M_{\rm env}^{\rm min} = M_H+\Delta M_{\rm DUP}$) and the surface
mass fractions of \chem{12}C, \chem{14}N and \chem{16}O at the time of pulse
ignition.

\begin{landscape}

\begin{deluxetable}{ccccccccccccc}
\tabletypesize{\scriptsize}
\tablecolumns{13} 
\tablewidth{0pc} 
\tablecaption{Selected Properties of the 7\msun\ AGB model}

\tablehead{
$M_H$  & $\Delta M_{\rm env}^{\rm H}$ &  $\Delta t_{\rm pul}$ &  $\Delta t_{\rm
inter}$ & $T_{\rm pul}^{\rm max}$
& $\rho_{\rm pul}^{\rm max}$ & $L_{\rm He}^{\rm max}$ & $\Delta M_{\rm pul}^{\rm H}$ &
$M_{\rm pul}$ & $\Delta M_{\rm DUP}$ & \multicolumn{3}{c}{Surface mass 
fraction} \\ 
\cline{11-13} \\
($M_{\odot}$) & ($10^{-3}$\msun) & (yr) & ($10^3$ yr) &  ($10^{8}$K) &
(cgs) & ($L_{\odot}$) & ($10^{-3}$\msun) &  ($10^{-3}$\msun)  &  ($10^{-3}$\msun) & \chem{12}C & \chem{14}N & \chem{16}O}

\startdata

1.0557 & 0.0800 & 2.25 & 0.8081 & 2.278 & 1297. & 7.244 (4) & 0.4341 & 0.263 & 0.0513 & 1.270 (-6) & 2.622E-05 & 9.519 (-8) \\
1.0559 & 0.0533 & 4.39 & 0.9787 & 2.367 & 1397. & 9.395 (4) & 0.2687 & 0.385 & 0.0391 & 1.103 (-6) & 2.649E-05 & 9.852 (-8) \\
1.0562 & 0.0444 & 4.97 & 1.0196 & 2.494 & 1459. & 1.437 (5) & 0.1955 & 0.484 & 0.0148 & 1.131 (-6) & 2.644E-05 & 1.041 (-7) \\
1.0565 & 0.0209 & 6.70 & 1.1066 & 2.565 & 1609. & 2.005 (5) & 0.1471 & 0.532 & 0.0069 & 1.179 (-6) & 2.636E-05 & 1.113 (-7) \\
1.0568 & 0.0179 & 5.56 & 1.0775 & 2.628 & 1599. & 2.537 (5) & 0.1305 & 0.543 & 0.0029 & 1.191 (-6) & 2.634E-05 & 1.195 (-7) \\
1.0571 & 0.0129 & 5.29 & 0.9974 & 2.674 & 1810. & 3.106 (5) & 0.1157 & 0.550 & -.0002 & 1.213 (-6) & 2.630E-05 & 1.271 (-7) \\
1.0574 & 0.0122 & 4.82 & 0.9392 & 2.783 & 1599. & 3.429 (5) & 0.1044 & 0.583 & -.0012 & 1.237 (-6) & 2.626E-05 & 1.335 (-7) \\
1.0577 & 0.0129 & 5.02 & 0.9436 & 2.747 & 1813. & 4.094 (5) & 0.0933 & 0.557 & 0.0001 & 1.251 (-6) & 2.623E-05 & 1.384 (-7) \\
1.0584 & 0.0174 & 4.05 & 1.2463 & 2.901 & 1824. & 7.483 (5) & 0.0707 & 0.600 & -.0029 & 1.277 (-6) & 2.619E-05 & 1.458 (-7) \\
1.0587 & 0.0165 & 4.33 & 1.2892 & 2.929 & 1869. & 8.776 (5) & 0.0600 & 0.600 & -.0018 & 1.307 (-6) & 2.614E-05 & 1.479 (-7) \\
1.0590 & 0.0167 & 4.14 & 1.3120 & 2.936 & 1940. & 9.615 (5) & 0.0574 & 0.586 & -.0034 & 1.314 (-6) & 2.613E-05 & 1.491 (-7) \\
1.0594 & 0.0149 & 4.50 & 1.3704 & 2.960 & 1734. & 9.870 (5) & 0.0546 & 0.567 & -.0050 & 1.330 (-6) & 2.611E-05 & 1.495 (-7) \\
1.0597 & 0.0149 & 3.87 & 1.3747 & 3.001 & 1784. & 1.222 (6) & 0.0509 & 0.578 & -.0042 & 1.342 (-6) & 2.609E-05 & 1.495 (-7) \\
1.0601 & 0.0153 & 4.34 & 1.4901 & 3.057 & 1961. & 1.581 (6) & 0.0446 & 0.602 & -.0028 & 1.356 (-6) & 2.607E-05 & 1.493 (-7) \\
1.0604 & 0.0142 & 4.33 & 1.7442 & 3.094 & 2745. & 2.116 (6) & 0.0375 & 0.669 & -.0833 & 1.388 (-6) & 2.602E-05 & 1.487 (-7) \\
1.0608 & 0.0147 & 4.11 & 2.0434 & 3.202 & 2822. & 3.848 (6) & 0.0292 & 0.771 & -.1142 & 1.561 (-6) & 2.876E-05 & 1.905 (-7) \\
1.0611 & 0.0147 & 4.11 & 2.5918 & 2.955 & 3055. & 6.657 (6) & 0.0250 & 0.878 & -.3037 & 1.910 (-6) & 3.515E-05 & 3.094 (-7) \\
1.0613 & 0.0121 & 4.46 & 2.3100 & 3.300 & 3019. & 6.254 (6) & 0.0238 & 0.815 & -.1720 & 3.186 (-6) & 5.996E-05 & 1.049 (-6) \\
1.0615 & 0.0118 & 4.12 & 2.2375 & 3.120 & 4184. & 2.763 (7) & 0.0155 & 0.518 & -.1289 & 3.981 (-6) & 7.567E-05 & 1.414 (-6) \\
1.0618 & 0.0111 & 4.59 &  ----- & 3.246 & 4238. & 9.356 (6) & 0.0193 & 0.897 &   ---- & 4.090 (-6) & 7.838E-05 & 1.067 (-6) \\

\enddata
\label{tabm7}
\end{deluxetable}
\end{landscape}

\begin{landscape}
\begin{deluxetable}{ccccccccccccc}
\tabletypesize{\scriptsize}
\tablecolumns{13} 
\tablewidth{0pc} 
\tablecaption{Selected Properties of the 3\msun\ AGB model}

\tablehead{
$M_H$  & $\Delta M_{\rm env}^{\rm H}$ &  $\Delta t_{\rm pul}$ &  $\Delta t_{\rm
inter}$ & $T_{\rm pul}^{\rm max}$
& $\rho_{\rm pul}^{\rm max}$ & $L_{\rm He}^{\rm max}$ & $\Delta M_{\rm pul}^{\rm H}$ &
$M_{\rm pul}$ & $\Delta M_{\rm DUP}$ & \multicolumn{3}{c}{Surface mass 
fraction} \\ 
\cline{11-13} \\
($M_{\odot}$) & ($10^{-3}$\msun) & (yr) & ($10^3$ yr) &  ($10^{8}$K) &
(cgs) & ($L_{\odot}$) & ($10^{-3}$\msun) &  ($10^{-3}$\msun)  &  ($10^{-3}$\msun) & \chem{12}C & \chem{14}N & \chem{16}O}

\startdata
\tablenotemark{a}\ \ 0.7576 &  21.6866 & 139.83 &  27.63   &   2.339 &   2209. & 3.227 (5) &   0.1494 &   5.237 &    2.8497 & 9.63 (-17) & 1.09 (-14) & 3.06 (-16) \\
\tablenotemark{a}\ \ 0.7604 &   8.4066 &  92.25 &  27.09   &   2.691 &   1821. & 5.926 (6) &   0.0707 &   5.502 &    1.0034 & 2.65 (-11) & 6.91 (-10) & 5.59 (-12) \\
\tablenotemark{a}\ \ 0.7619 &  13.2376 & 150.31 &  15.18   &   2.513 &   2389. & 6.251 (5) &   0.0612 &   4.659 &    1.0784 & 5.69 (-11) & 2.247 (-9) & 2.29 (-11) \\
\tablenotemark{a}\ \ 0.7653 &  10.1281 &  65.31 &  30.03   &   2.573 &   1065. & 1.628 (6) &   0.0819 &   5.311 &    2.2147 & 4.95 (-10) & 1.814 (-8) & 1.92 (-10) \\
\hline
0.7683 &   2.4285 &  66.12 &  30.98   &   2.802 &   1780. & 1.208 (7) &   0.0972 &   6.045 &   -0.0205 & 2.071 (-7) & 9.312 (-8) & 1.846 (-6) \\
0.7713 &   2.1531 &  31.00 &  30.81   &   2.867 &   1780. & 1.568 (7) &   0.0684 &   5.986 &   -0.0351 & 2.070 (-7) & 9.367 (-8) & 1.845 (-6) \\
0.7742 &   1.9328 &  26.61 &  30.75   &   2.913 &   1896. & 1.773 (7) &   0.0656 &   5.696 &   -0.0378 & 2.070 (-7) & 9.408 (-8) & 1.845 (-6) \\
0.7772 &   1.7731 &  24.67 &  29.40   &   2.933 &   1804. & 2.270 (7) &   0.0545 &   5.540 &   -0.0433 & 2.069 (-7) & 9.447 (-8) & 1.845 (-6) \\
0.7802 &   1.6224 &  21.39 &  28.32   &   2.977 &   1951. & 2.407 (7) &   0.0454 &   5.290 &   -0.0454 & 2.069 (-7) & 9.484 (-8) & 1.844 (-6) \\
0.7830 &   1.5284 &  21.96 &  27.39   &   2.984 &   1937. & 2.367 (7) &   0.0522 &   5.051 &   -0.0476 & 2.068 (-7) & 9.519 (-8) & 1.844 (-6) \\
0.7857 &   1.4341 &  21.38 &  26.45   &   2.989 &   1830. & 2.503 (7) &   0.0485 &   4.888 &   -0.0485 & 2.068 (-7) & 9.554 (-8) & 1.844 (-6) \\
0.7884 &   1.3783 &  17.46 &  27.81   &   2.984 &   1774. & 2.381 (7) &   0.0593 &   4.723 &   -0.0473 & 2.068 (-7) & 9.588 (-8) & 1.843 (-6) \\
0.7912 &   1.3000 &  33.41 &  35.25   &   3.027 &   1851. & 2.827 (7) &   0.0519 &   4.702 &   -0.0641 & 2.067 (-7) & 9.621 (-8) & 1.843 (-6) \\
0.7944 &   0.8771 &  26.09 &  42.44   &   3.137 &   1930. & 6.745 (7) &   0.0133 &   5.150 &   -0.3435 & 1.299 (-6) & 2.402 (-7) & 3.836 (-6) \\
0.7983 &   0.3438 &  26.79 &  45.46   &   3.278 &   2142. & 2.256 (8) &   0.0118 &   6.027 &   -1.0055 & 2.274 (-5) & 3.122 (-7) & 2.333 (-5) \\
0.8020 &   0.1770 &  39.25 &  47.20   &   3.412 &   2566. & 4.126 (8) &   0.0173 &   6.446 &   -1.4912 & 1.245 (-4) & 3.152 (-7) & 4.265 (-5) \\
0.8054 &   0.1085 &  34.24 &  48.09   &   3.469 &   2653. & 6.075 (8) &   0.0171 &   6.624 &   -1.8280 & 2.949 (-4) & 3.249 (-7) & 5.269 (-5) \\
0.8087 &   0.0705 &  49.59 &  48.51   &   3.564 &   3090. & 7.975 (8) &   0.0164 &   6.725 &   -2.1434 & 5.091 (-4) & 3.662 (-7) & 5.875 (-5) \\
0.8118 &   0.0781 &  61.04 &  55.31   &   3.612 &   3186. & 1.045 (9) &   0.0111 &   6.806 &   -2.3621 & 7.081 (-4) & 1.269 (-5) & 6.474 (-5) \\
0.8150 &   0.0697 &  57.29 & 122.18   &   3.683 &   3371. & 1.591 (9) &   0.0112 &   7.088 &   -3.4720 & 3.712 (-4) & 6.690 (-4) & 7.148 (-5) \\
0.8190 &   0.0120 &  87.99 & 150.48   &   3.908 &   3999. & 7.121 (9) &   0.0135 &   8.735 &   -4.9227 & 5.190 (-5) & 1.670 (-3) & 7.992 (-5) \\
0.8226 &   0.0101 &  65.82 & 130.23   &   3.956 &   3832. & 1.496 (10) &  0.0082 &   9.586 &   -7.1713 & 7.372 (-5) & 2.400 (-3) & 9.216 (-5) \\
0.8240 &   0.0074 &  66.75 &   -----  &   4.054 &   4443. & 1.753 (10) &  0.0111 &   9.729 &   -----   & 1.031 (-4) & 3.261 (-3) & 1.046 (-4) \\

\enddata
\tablenotetext{a}{Carbon injection episodes}
\label{tabm3}
\end{deluxetable}
\end{landscape}


We also find that after the disappearance of the HCS and before the
envelope penetrates the region polluted by the HCS, a secondary convective zone
can develop at the border of the H rich region. This feature, illustrated
in Fig. \ref{m2ingest} in the 2\msun\ model, depends on the mesh resolution
and numerical scheme (see below). Although this secondary convective
shell never mixes with the underlying \chem{12}C and \chem{4}He rich
layers, this situation is unstable with respect to the Schwarzschild
criterion because of the large positive value of $\nabla_{\mathrm{rad}} -
\nabla_{\mathrm{ad}}$ at the bottom of the convective region
(e.g. Castellani et al. 1971a, 1971b). When $\nabla_{\mathrm{rad}} > \nabla_{\mathrm{ad}}$,
any perturbation of the convective boundary would give rise to extra
mixing, and in particular carbon could be injected into the convective zone
with the possible triggering of a new H flash and further chemical
contamination of the envelope. The formation of a ``secondary
convective shell'' above the H discontinuity has also been found by Dominguez
et al. (2000), after 9 weak instabilities during the evolution of a 5\msun\
primordial star, and it strikingly resembles the one depicted here 
except that in their simulation, it takes place before the C-injection
(see their Fig. 7). Finally note that in our simulations this feature, if
real, has no influence on the subsequent evolution since it does not mix
with the underlying C-rich layers.

The existence of C-injection episodes at the beginning of the AGB phase has
recently been found by CDLS in their computations of zero-metal
intermediate-mass stars. In these computations, this process follows the
same course of events : a convective shell develops at the base 
of the H rich layer, ingests carbon and initiates an H flash. The
additional energy released by this H flash subsequently leads to a deeper
penetration of the convective envelope which reaches the CNO enriched HCS
and pollutes the envelope. Our results thus confirm these independent
computations and the unusual circumstances by which primordial stars enter
the TP-AGB phase.

Finally, we would like to discuss briefly the sensitivity of the results to
the numerical treatment. During the penetration of the He-driven convective
zone into the H rich layers and subsequent carbon injection event, we
performed numerous tests, adopting different time steps and different mass
zoning. In these tests, we also accounted or not for
a small  overshooting characterized by $d=0.05H_P$. Our numerical
experiments indicate that when  extra-mixing at the
borders of the convective shells is allowed, the penetration of the HCS in the
underlaying C-rich is present in all the models, irrespective of the mesh
point and time-stepping. However, when overshooting is not included, the
C-injection then depends on the spatial and temporal resolution. 
We also report that in some cases a secondary convective zone develops just
above the H-He discontinuity, as previously reported and shown in
Fig. \ref{m2ingest}. The 
appearance of this feature is however arbitrarily dependent on the adopted
temporal and spatial resolutions. The amplitude of the H flash and the
extent of the HCS vary slightly from one computation to the other. In
particular, adopting a larger timestep leads in general to a larger value
of \lH\ at its maximum which is the opposite effect of having a more
refined mesh. As a consequence of these numerical uncertainties, the
envelope chemical composition after the dredge-up event may vary by a
factor of a few. The stronger the H flash, the deeper the convective
envelope plunges and the higher is the surface metallicity. While these
differences appear to be relatively large, we have to remember that the
envelope composition of the stars that experience these C-injections is
almost devoid of metals. Therefore, a small difference in the dredge- up
mass can lead to significant changes after the mixing events. Basically, it
is much easier to raise the metallicity from almost zero to $\sim 10^{-6}$
than from $10^{-6}$ to $10^{-5}$. The important point for the subsequent
evolution is that a substantial amount of carbon has been deposited into
the envelope.  Remembering that an envelope mass fraction of
\chem{12}C as low as $X_{12\mathrm{C}} \simeq 10^{-7}$ is sufficient to
switch H burning to the CNO cycle, we realize that an uncertainty in the
amount of C dredged-up in the envelope by as much as an order of magnitude
will not result in a change in the subsequent evolution, because the
C-injection has already raised the \chem{12}C abundance well above this 
limit. The influence on the yields will also be negligible since the main
chemical enrichment of the surface comes from the third dredge-up event
(3DUP). As a conclusion, even though the modeling of this mixing phase
suffers from some numerical uncertainties, its occurrence is well
established from our simulations and those of CDLS and we are confident
that the exact modeling of this process do not influence the subsequent
evolution of the thermally pulsing AGB stars substantially.

\subsubsection{The degenerate pulses} 
\label{m5}

Our 4 and 5\msun\ models exhibit a different kind of thermal pulses which
we refer to as ``degenerate pulses'' following the designation given by
Frost et al. (1998). We illustrate their characteristics by returning to
the evolution of our 5\msun\ model.

After a series of 12 weak instabilities, a convective pulse
develops and the 5\msun\ model undergoes the carbon injection phenomenon
described previously. After $\sim 15800$~yr, another  
standard thermal pulse takes place.  Because of the relatively high
degeneracy in the He shell, this second pulse ignites very deeply (at $M_r
\simeq 0.901$\msun) and gives rise to a very extended convective
region. The helium generated luminosity reaches $\LHe \simeq 1.5\times
10^7\Lsun$ at its peak and is maintained above $10^3$\lsun\ for $\approx$
2900~yr which allows for an efficient He depletion in the intershell.

Following the pulse, the convective envelope deepens and produces a
dredge-up characterized by $\lambda \sim 0.4$, where $\lambda = \Delta
M_{\mathrm{DUP}}/\Delta M_{\mathrm{H}}$ represents the ratio of the mass
which has been dredge-up ($\Delta M_{\mathrm{DUP}}$) to the mass advance of
the HBS between the last 3DUP and the beginning of the pulse ($\Delta
M_{\mathrm{H}}$). During the subsequent interpulse, hot bottom burning
starts operating and leads temporarily to the production of \chem{7}Li due
to the Cameron-Fowler (1971) mechanism. It also increases the surface
abundances of \chem{13}C and \chem{14}N at the expense of \chem{12}C.
After $\simeq 14 000$~yr, a stronger instability is triggered in the HeBS
($\LHe \simeq 5\times 10^7$\lsun) and a deep third dredge-up ensues
($\lambda \sim 0.7$), which rapidly extinguishes the instability.  This
quick deactivation of the third pulse (in $\sim 10$\,yrs) does not allow
for a large depletion of the helium present in the intershell convective
tongue and a long tail of unburnt He ($Y \la 0.1$) is left behind the pulse
(Fig. \ref{degenp} lower panel).
When this He tail reignites after $\sim 70$~yr, a second convective
instability develops (Fig. \ref{degen}). It lasts for $\sim 40$~yrs and
generates a nuclear luminosity of $\LHe \sim 10^6$\lsun. At the end of this
secondary pulse, the He tail has disappeared and a large fraction of the He
present in the intershell has also been burnt (Fig. \ref{degenp} upper
panel). We also found a degenerate pulse in our 4\msun\ model
(Fig. \ref{degen}). Its characteristics are very similar: the maximum He
generated luminosities during each instability are $\LHe \simeq 8\times
10^6$ and $8\times 10^5$\lsun, respectively, slightly lower than in the
5\msun\ model and the duration of the two convective pulses, $\sim 30$ and
60~yrs. We note however that in the 4\msun\ case, the two instabilities are
separated in time by only $\sim 10$~yr and look like a failed pulse.

This phenomenon is very similar to the one described in detail by Frost et
al. (1998), who first reported the appearance of the so called
``degenerate'' thermal pulses in their computation of a low metallicity
($Z=0.004$) 5\msun\ AGB model. The occurrence of these instabilities
results from the fact that the star experiences a deep dredge-up that
quenches rapidly the instability, preventing an efficient He depletion in
the intershell. It is also favored by a higher degeneracy which
prevents efficient helium burning because of the lower temperature found at
the base of the pulse. We present in Fig. \ref{eta} the evolution of the
degeneracy parameter $\eta$ and temperature, before and after the standard
pulse preceding the degenerate instability, during the He accretion phase
and just before the ignition of the (degenerate) pulse. As shown, the
degeneracy is lifted during the pulse and the temperature increases
in the intershell region (solid and dotted lines). During the He accretion
phase, $\eta$ increases and the temperature excess dissipates (short-dashed
line). Finally, when the degenerate pulse develops, the parameter $\eta$ is
larger and the temperature lower than in the previous pulse.
However, contrary to Frost et al. (1998), our degenerate pulse develops
very early in the thermally pulsing AGB phase, probably as the result of
the higher degeneracy of our zero metallicity stars. We also note that the
amplitude of the generated luminosity and amount of dredge-up before the
pulse are smaller in our simulations. These calculations are totally
independent of those of Frost et al. (1998) and they verify that degenerate
pulses indeed occur (at least in models, if not in real stars !).

After this series of events, our 4 and 5\msun\ stars enter a ``standard''
AGB evolution phase. We note however, that the temperature and degeneracy
keep increasing at the base of the helium shell.  This situation is
favorable for a later development of a degenerate pulse (see also Frost et
al. 1998).  Finally, note that CDLS do not report any
``strange'' behavior such as the one described here in the evolution of
their 5\msun\ $Z=0$ model.

\subsubsection{Nucleosynthesis}

When the temperature in the convective pulse exceeds $\sim 3 \times 10^8$K,
as it is found in some late thermal pulses or in more massive stars like in
the 5\msun\ model, a large number of neutrons is released via the reaction
\chem{22}Ne($\alpha$,n)\chem{25}Mg. As an example, during the 17th
pulse of our 3\msun\ model,  at the peak of neutron production, the neutron
density $N_n > 3 \times 10^{10} 
\,{\mathrm{cm}}^{-3}$ everywhere in the pulse and reaches a value of $N_n
\simeq  10^{13} \,{\mathrm{cm}}^{-3}$ at the base of the convective
instability where the temperature $\simeq 3.9 \times 10^8$K.  At lower
temperatures, a marginal source of neutrons is provided by the
\chem{13}C($\alpha$,n)\chem{16}O chain of reaction, the \chem{13}C coming
initially from the ashes of the HBS.  These neutrons mainly react with the
\chem{14}N and some \chem{14}C is produced, which in turn, reacts with
$\alpha$ particles leading to \chem{18}O production and then by another
$\alpha$ capture to the formation of \chem{22}Ne.  The \chem{22}Ne is also
subject to ($\alpha,\gamma$) and ($\alpha$,n) reactions and the abundance
of \chem{26}Mg and \chem{25}Mg are increased, respectively. Note that
after the rapid depletion of \chem{14}N, some  \chem{13}C is produced by
\chem{12}C(n,$\gamma$) (see Tab. \ref{chempulse}).

Further neutron captures on \chem{22}Ne and \chem{26}Mg increases the
\chem{23}Na and \chem{27}Al abundances respectively, and after the
bottle-neck formed by \chem{33}S(n,$\alpha$)\chem{30}Si is passed, neutron
captures can proceed on heavier elements (see Goriely \& Siess 2001 for
more details). Some \chem{19}F is also synthesized by
\chem{18}O(p,$\alpha$)\chem{15}N($\alpha,\gamma$)\chem{19}F. In the inner
region, \chem{12}C continues to be produced via the \3a reactions and it is
partly destroyed by $\alpha$ capture to produce \chem{16}O.  In the upper
part of the pulse, the largest nuclear fluxes result from proton captures
on \chem{12}C, \chem{14}N, \chem{16}O and \chem{18}O, the protons mainly
supplied by \chem{14}N(n,p)\chem{14}C. We also found that the chain of
\chem{16}O(n,$\gamma$)\chem{17}O($\alpha$,n)\chem{20}Ne contributes to the
build-up of \chem{20}Ne, while restoring a substantial number of neutrons.
The abundances of some key elements during the 17th pulse of our 3\msun\
model are presented in Tab. \ref{chempulse}

\begin{deluxetable}{ccccccccccc}
\tablecolumns{11}
\tabletypesize{\scriptsize}
\tablewidth{0pc} 
\tablecaption{Mass fractions of some elements inside selected thermal pulse of the
  3\msun\ model. The abundances are shown near the beginning and end of the
convective pulse.}
\tablehead{pulse\tablenotemark{a} & \chem{13}C & \chem{14}C & \chem{14}N & \chem{22}Ne & \chem{23}Na
  & \chem{25}Mg & \chem{26}Mg & \chem{27}Al & \chem{30}Si & \chem{36}S}
\startdata
2 & 1.26(-9) & 2.05(-7) & 6.43(-6) & 1.30(-2) & 8.80(-5) & 1.36(-4) & 4.08(-4) & 9.16(-4) & 3.55(-3) & 5.32(-4) \\
2 & 4.36(-9) & 3.37(-8) & 5.1(-13) & 7.36(-3) & 4.66(-5) & 1.43(-4) & 2.80(-4) & 4.63(-4) & 1.96(-3) & 2.67(-4) \\

10 & 1.7(-12) & 1.43(-6) & 7.21(-5) & 8.02(-4) & 5.66(-6) & 8.45(-6) & 1.80(-5) & 3.21(-5) & 3.92(-7) & 6.69(-7) \\
10 & 3.4(-11) & 3.63(-5) & 1.30(-9) & 9.40(-4) & 1.79(-5) & 2.38(-5) & 6.47(-5) & 1.31(-4) & 1.85(-6) & 1.61(-6) \\

17 & 4.7(-11) & 8.35(-7) & 1.98(-4) & 4.30(-4) & 4.16(-6) & 3.04(-5) & 1.05(-4) & 3.91(-6) & 4.75(-7) & 1.37(-8) \\
17 & 1.4(-12) & 9.9(-11) & 1.9(-13) & 9.27(-4) & 3.22(-5) & 3.55(-4) & 1.37(-3) & 2.82(-5) & 2.76(-6) & 7.49(-8) \\ 
\enddata
\tablenotetext{a}{The pulse number is counted after the end of the
carbon injection episodes. Details about the evolution of this model are 
provided in Tab.\ref{tabm3}}
\label{chempulse}
\end{deluxetable}

Finally, the main reactions which take place in the HCS are the CNO
reactions. The fact that the elements synthesized during the pulse are
mixed in the HCS at relatively high temperatures also leads, but to a much
lesser extent, to marginal $\alpha$-capture on CNO isotopes. However,
because of the small abundance of Ne and Mg in the HCS, the NeNa and MgAl
cycles are not working efficiently.

\subsubsection{The occurrence of dredge-up episodes}
\label{dup}




Third dredge-up episodes are responsible for the formation of carbon
stars (see e.g. Iben \& Renzini 1982) and are also necessary (but not
sufficient) to account for the synthesis and observations of s-elements in
AGB stars. However, details of 3DUP are dependent on numerous
factors such as the numerical scheme (e.g. Frost \& Lattanzio 1996), the
spatial or temporal resolution of the simulations (Straniero et al. 1997),
the input physics (choice of opacity, mass loss rate) and is favored by the
presence of extra-mixing or overshooting. In past years, new formalisms
for overshooting (e.g. Herwig et al. 1997) and rotationally-induced mixing
(Langer et al. 1999) have been developed in order to facilitate the
occurrence of 3DUP to address the issues related to the luminosity function
of carbon stars (so called C-star mystery) and production of s-elements.

The 3DUP is characterized by the penetration of the convective envelope into
a region of significantly different chemical composition and as a
consequence a discontinuity in $\nabla_{\mathrm{rad}} -
\nabla_{\mathrm{ad}}$ appears at the inner convective boundary. As
described previously, this situation is unstable and favors convective
overshoot from the envelope into the underlaying layers. To account for
this natural tendency of the convective envelope to cross the strict limit
defined by the Schwarzschild criterion, we have allowed for a small
overshooting characterized by $d = 0.05 H_P$. In this context, we were able
to reproduce the third dredge-up (3DUP) and by the end of our computations,
it is present in all of our models with $1.5 \la M \la 7$\msun. It starts
operating at the first (degenerate) pulse of our 4 and 5\msun\ stars and is
characterized by relatively large initial values of $\lambda \ge 0.7$ . In
the 2, 3 and 7\msun\ models, the 3DUP comes into play at the 6$^{th}$,
9$^{th}$ and 15$^{th}$ pulse, respectively (Figs. \ref{m2kip}, \ref{m3kip}
and \ref{m7kip}). Note that CDLS also find 3DUP events in their 7\msun\
model after 13$th$ or 14$th$ pulses, depending on their prescription for
extra-mixing ($\beta = 0.005$ and 0.01, respectively). At the termination
of our computations, the 1\msun\ model has undergone two deep mixing events
during the 4$^{th}$ and 8$^{th}$ pulses (Fig. \ref{m1kip}) but we suspect
that these may be artifacts due to a smaller spatial resolution at that
time. Finally, we report that in the 1.5\msun\ model, the 3DUP follows the
first convective instability. The evolution of the latter star is peculiar
in the sense that it develops extremely strong pulses ($\LHe \sim 10^8 -
10^9$\lsun) since the beginning of the TP-AGB phase. This behavior is
certainly related to the higher degeneracy of the He layers in this model,
which  mass  lies close to the
upper mass limit for degenerate He core ignition. As a consequence of the
higher degeneracy, the pulses are stronger and the convective penetration
deeper. Inspection of the structure indicates that the extent of the
convective instabilities is much larger than in the two adjacent mass
tracks (1 and 2 \msun) and
also that the pulse ignites near the base of the convective instability.

The impact of the 3DUP on the surface abundances is discussed in 
$\S$ \ref{nucevol}.

\section{Chemical evolutionary consequences}
\label{nucevol}

One of the great interest in the first generation of stars is their effect
on the chemical evolution of the universe. We have seen that these stars
do, indeed, experience the third dredge-up on the AGB, and also hot bottom
burning. The consequences of these events for the chemical history of the
universe are our concern in this section.

The largest uncertainties in calculating the chemical yields are the
unknown mass-loss history of the stars and the amount of dredge-up material
which affects both the chemical composition and the efficiency at which
stellar matter is returned to the interstellar medium. If, as is believed,
the mass-loss in low- and intermediate-mass stars is driven by (or at least
strongly coupled to) radiation pressure on grains, then an envelope free
from grains could have a very reduced mass-loss. However, if the
third dredge-up takes place, as found in these simulations, the envelope is
enriched in metals (mainly C, N and O) and the mass loss is switched on,
although later in the evolution compared to higher metallicity stars. Thus
the initial-final mass relation is likely to favor the production of more
massive white dwarfs at $Z=0$ than for other values of metallicity. On the
other hand, the higher effective temperatures characteristic of these stars
might make it more difficult for grains to form (even though nova systems
show that grains form even in unlikely environments). As a consequence,
the uncertainty in the form of both the mass-loss rate and the efficiency
of the third dredge-up are still a major limitation at
present and prevents us from calculating yields for our models.

Prior to the AGB phase, the only significant consequence of the (second)
dredge-up is the increase in helium abundance. But as we have noted above,
all of our models do experience the third dredge-up. This is a major source
of \chem{12}C, of course, but also of other species. For example, the
\chem{12}C is processed by the hydrogen burning shell and the CN cycle thus
produces \chem{14}N. When this \chem{14}N is exposed to high temperatures
and a significant flux of $\alpha$ particles during the next thermal pulse,
it is completely burned into \chem{22}Ne. The \chem{22}Ne is, in turn,
partially burned by further $\alpha$-captures into \chem{25}Mg (if subject
to ($\alpha$,n) reactions) or \chem{26}Mg (if subject to ($\alpha,\gamma$)
instead). Details depend on the stellar mass, of course, as the main
parameter determining the shell temperatures, but we expect both to occur
to some extent.

We have also seen that for masses above about 2\msun\ there is activation
of hot bottom burning. This is in distinct contrast to the standard
assumption in the literature that the minimum mass for HBB is between 4 and
5\msun. Although this is true for solar metallicity, a decrease in $Z$
will cause the onset of HBB to occur at lower masses, as found in Lattanzio
et al. (2001). Clearly these stars will be significant sources of
\chem{14}N in the early Universe (see top panel of Figs. \ref{m2kip},
\ref{m3kip}, \ref{m7kip} and \ref{m1kip}).

But HBB does more than burn \chem{12}C into \chem{14}N. It will also
process the \chem{22}Ne into \chem{23}Na, as well as the \chem{25}Mg and
\chem{26}Mg into \chem{26}Al and \chem{27}Al, respectively.  A potentially
important reaction is the production of \chem{7}Li via the Cameron-Fowler
mechanism. But although \chem{7}Li is rapidly and abundantly produced by
HBB it is also destroyed on subsequent passes through the bottom of the
convective envelope. Hence the Li-rich phase is only temporary.  Whether
these stars can add to the initial Li abundance in the early universe will
depend crucially on the mass-loss involved: if the mass-loss is significant
while the surface Li abundance is high, then they will produce a large Li
yield. If, however, the Li is depleted before the high mass-loss rates
begin (a situation believed to occur in the current generation of stars)
then the yields will be negligible, or even negative.  The reader is
referred to Travaglio et al. (2001) for a fuller discussion.

Figures \ref{m2kip}, \ref{m3kip},  \ref{m7kip}  and \ref{m1kip} depict the evolution of
the structure concomitantly to the evolution of the surface abundances of
some keys elements. As can be seen, during the initial carbon ingestion, the
stars soon become N-rich as a consequence of dredging-up to the
surface the products of CNO cycle which was operating in the HCS. Following
a few thermal pulses, the 3DUP takes place and the abundances of C, O, Mg,
Ne and Na increase.
Note that the stars rapidly become C-rich. However it is not clear
whether primordial stars will maintain their large C/O ratio or
not. Indeed, the evolution of the C/O ratio depends on the competition
between 3DUP (which increases the C abundance) and HBB (which destroys it)
which in turn depends on the envelope mass, i.e. on mass loss. Frost et
al. (1998) showed that, as the envelope mass decreases, HBB shuts down
first while 3DUP continues.  As far as our computations are concerned, our
low- and intermediate-mass stars show a C/O ratio larger than one, but for
how long this feature will be maintained is presently unknown. When 
the temperatures at the base of the convective envelope approaches
$40\times 10^6$K, HBB is activated.  In the 3\msun\ and 7\msun\ models, the
signatures of this efficient nuclear burning 
is well illustrated : the \chem{12}C is rapidly converted into \chem{14}N
which becomes the most abundant metal in the envelope and the \chem{22}Ne
after proton capture produces a large amount of \chem{23}Na. However,
because of the maintenance of a high temperature at the 
base of the convective envelope ($T \ga 10^8$K, $\rho > 10 \gcm$) of the 7\msun\ model,
proton captures on \chem{23}Na destroy this element at the benefit of
\chem{20}Ne and \chem{24}Mg.

Finally, if mixing of protons is taking place at the base of the convective
envelope (e.g. due to diffusive overshooting) the protons will react on the
abundant \chem{12}C and form a ``\chem{13}C pocket'' which will be
responsible for a large release of neutrons (Iben \& Renzini 1982, Busso
et al. 1999 for a review). The neutrons will then be captured by the most
abundant nuclei, i.e. C,N,O and Ne and lead to s-process enrichment {\sl
despite} the absence of iron (Goriely \& Siess 2001). As a consequence, we
can expect s-process elements to have formed in primordial stars.

\subsection{Conclusions}

The evolution of low- and intermediate mass stars in the mass range $1\la M
\le 5$\msun\ is characterized by the development of mixing episodes at the
beginning of the AGB phase. During these events, a secondary 
convective shell develops at the H-He discontinuity, it expands and
overlaps with the underlying carbon rich layers left by the receding pulse. 
The engulfed carbon initiates a H flash which further expands the structure and
produces a deeper penetration of the envelope. The envelope eventually
reaches the region previously occupied by this secondary shell and CNO
catalysts are dredged-up to the surface by the carbon injection mechanism.

This mechanism then allows low-
and intermediate-mass stars to successfully achieve the thermally pulsating
phase. These results confirm recent similar findings by CDLS.  Assuming a
small degree of overshooting at the base of the convective envelope, we find that a 
third dredge-up rapidly takes place and that hot bottom burning is already
active in stars with $M \ga 2$\msun. These processes contribute to
significant surface enrichment in \chem{14}N (which is further enhanced by
HBB), \chem{16}O, \chem{12}C (but the latter is partially destroyed by HBB) and
\chem{4}He with possible Mg pollution. Primordial AGB stars are also found
to be a potential site for s-process nucleosynthesis, despite the absence
of iron seeds (Goriely \& Siess 2001, Busso et al. 2001)

The metal enrichment of the envelope activates mass loss,
although later in the evolution compared to higher metallicity stars. As a
consequence, we expect primordial stars to form more massive white dwarf
cores than their more metal rich counterparts, but large uncertainties still
remain on the exact form of the mass loss history.

The first generation of stars has important cosmological consequences. In
particular, recent works have shown that adopting zero metallicity models
instead of the cooler very metal poor ones, can significantly modify the
reionization epoch (e.g. Cojazzi et al. 2000, Tumlinson \& Shull 2000,
Bromm et al. 2001). The formation of primordial stellar remnants may also
account for some fraction of the Galactic dark matter (e.g. Chabrier 1999).
The detection of carbon
(e.g. Songaila \& Cowie 1997) and nitrogen (e.g. Lu al. 1998) enrichments,
as well as of metal lines, in the intergalactic medium at high redshifts,
suggests primordial AGB stars as strong candidates for the origin of some
of these elements (Abia et al. 2001). In the same vein, observations of
chemically peculiar metal-poor stars are difficult to explain in the
framework of current mixing mechanisms (e.g. Denissenkov et
al. 1998). Whence the idea of an evolutionary scenario in which a first
generation of stars pollutes the proto-globular cluster cloud, producing
initial abundance anomalies which survive and are observed nowadays in
population II stars.  These examples illustrate only a few of the relevant 
issues associated with the unique evolutionary properties of
the first stars. A more extended study, covering an even larger parameter
space than the present work, 
as well as a better understanding of some of the physical processes are
needed to address all of these issues.

\acknowledgements

The authors gratefully acknowledge valuable comments by an anonymous
referee. These comments have significantly improved the presentation of this
paper.  LS wishes to thanks the STScI and Monash Department of Mathematics
and Statistics for their hospitality during his successive visits. This
work has been funded in part by the grant 938-COS191 from the John
Templeton Foundation in part by the Australian Research Council and in part
by the French-Australian International Program for Scientific Collaboration
``Stellar Populations, Dark Matter and Galaxy Surveys". LS also acknowledges
support from a European TMR ``Marie Curie'' fellowship at ULB.

\newpage

\figcaption{Central carbon abundance as a function of central H abundance (in
mass fraction). From right to left the different curves correspond to
stellar masses equal to 0.8, 1.0, 1.5, 2, 3, 4, 5, 7, 10, 15 and 20\msun,
respectively. The ignition of the CNO cycle during the main sequence occurs
when $\log (^{12}\mathrm{C}) \simeq -11.5$ and can be identified by the
bump in the different curves, which corresponds to the appearance of a
convective core.}
\label{XC}

\figcaption{HRD diagram for the complete set of $Z=0$ evolutionary tracks,
from 0.8 to 20\msun. The dots and stars on the tracks indicate the end of
the H and He core burning phases, respectively. The 0.8 and 1.0\msun\
experience the first dredge-up and at the tip of their red giant branch
undergo the He flash (triangle in case of the 1\msun\ model).}
\label{hrd}

\figcaption[]{Evolution of the convective regions (lower panel) and of the
  nuclear luminosity (upper panel) of the 1\msun\ model during the He core
  flash.  Convective regions are represented by hatched areas.}
\label{flash}

\figcaption[]{Evolution of the structure of the 5\msun\ model. The hatched
areas correspond to a convective region, the heavy solid lines delineates the
H and He burning regions where $\varepsilon_{\mathrm{nuc}} > 5\,
\mathrm{erg\,g\,s^{-1}}$.}
\label{kippen5}

\figcaption{Structural evolution of the 3\msun\ model during the carbon
injection (lower panel). The convective regions are hatched and the dashed
lines delineate the H and He burning regions. The dotted line inside the
burning regions corresponds to the locus of maximum nuclear energy
production. The top panel shows the evolution of the H (solid) and He
(dotted line) generated luminosities.}
\label{m3ingest}

\figcaption{Same as Fig. \ref{m3ingest} but for our 2\msun\ model}
\label{m2ingest}

\figcaption{Abundance profiles in mass fraction of \chem{4}He (solid line),
\chem{12}C (dotted line) and \chem{16}O (dashed line) during the degenerate
pulse of our 5\msun\ model. In the lower and upper panels, the thick and
thin lines correspond to the beginning (b) and the end (e) of the pulses
and are separated in time by 10 and 40~yrs, respectively.  Note the large
amount of He remaining in the intershell and long tail of unburnt He left
after the first pulse.}
\label {degenp}

\figcaption{Convective zones as a function of model number for the degenerate
pulses in the 4 and 5\msun\ models.}  
\label{degen}

\figcaption{Evolution of the profiles of temperature and degeneracy parameter
$\eta$ in the 5\msun\ model. The solid and dotted lines correspond to the
beginning and end of the pulse preceding the degenerate pulse,
respectively. The short- and long-dashed depict the situation during  the
He accretion phase and at the beginning of the degenerate pulse,
respectively. Note the higher degeneracy and lower temperature at the
ignition of this latter pulse.}
\label{eta}

\figcaption{The evolution of the structure of the 2\msun\ model is presented
in the lower panel. The dotted region represents the convective envelope
and the dashed lines, the locus of maximum energy generation by the HBS and
HeBS. In the upper panel, the surface abundances (in mass fraction) of
\chem{12}C, \chem{16}O, \chem{14}N, \chem{22}Ne, \chem{23}Na, \chem{26}Mg
and \chem{25}Mg are also represented.}
\label{m2kip}

\figcaption{Same as Fig. \ref{m2kip} but for our 3\msun\ model. Note the large
production of \chem{14}N and \chem{23}Na due to efficient HBB.}
\label{m3kip}

\figcaption{Same as Fig. \ref{m2kip} but for our 7\msun\ model. Note the rapid
depletion of \chem{23}Na by proton captures due to the high temperature at
the base of the convective envelope ($T \sim 10^8$K).} 
\label{m7kip}

\figcaption{Same as Fig. \ref{m2kip} but for our 1\msun\ model. Note that
dredge-up only occurs in the 4th and 8th pulse.}
\label{m1kip}

\end{document}